\documentclass[twocolumn,showpacs,floatfix,prb]{revtex4}

\usepackage{graphicx}
\usepackage{dcolumn}

\begin{document}

\input epsf.sty

\draft
\widetext

\title
{
Stripe order, depinning, and fluctuations
in La$_{1.875}$Ba$_{0.125}$CuO$_{4}$ and
La$_{1.875}$Ba$_{0.075}$Sr$_{0.050}$CuO$_{4}$
}


\author{M. Fujita}\email{fujita@imr.tohoku.ac.jp}
\author{H. Goka}
\author{K. Yamada}

\affiliation{
Institute for Materials Research, Tohoku University, Katahira, Sendai 
980-8577, Japan
}

\author{J. M. Tranquada}

\affiliation{
Physics Department, Brookhaven National Laboratory, Upton, NY 11973
}

\author{L. P. Regnault}

\affiliation{
CEA/Grenoble, D\'{e}partement de Recherche Fondamentale sur la
Mati\'{e}re Condens\'{e}e, 38054 Grenoble cedex 9, France
}

\date{\today}



\begin{abstract}
We present a neutron scattering study of stripe correlations measured on
a single crystal of La$_{1.875}$Ba$_{0.125}$CuO$_{4}$.  Within the
low-temperature-tetragonal (LTT) phase, superlattice peaks indicative of
spin and charge stripe order are observed below 50 K.  For excitation
energies $\hbar\omega\le12$~meV, we have characterized the magnetic
excitations that emerge from the incommensurate magnetic superlattice
peaks.  In the ordered state, these excitations are similar to spin
waves.  Following these excitations as a function of temperature, we find
that there is relatively little change in the {\bf Q}-integrated
dynamical spin susceptibility for $\hbar\omega\sim10$~meV as stripe
order disappears and then as the structure transforms from LTT to the
low-temperature-orthorhombic (LTO) phase.  The {\bf Q}-integrated signal
at lower energies changes more dramatically through these transitions, as
it must in a transformation from an ordered to a disordered state.  We
argue that the continuous evolution through the transitions provides
direct evidence that the incommensurate spin excitations in the
disordered state are an indicator of dynamical charge stripes.  An
interesting feature of the thermal evolution is a variation in the
incommensurability of the magnetic scattering.  Similar behavior is
observed in measurements on a single crystal of
La$_{1.875}$Ba$_{0.075}$Sr$_{0.050}$CuO$_{4}$; maps of
the scattered intensity in a region centered on the antiferromagnetic
wave vector and measured at $\hbar\omega=4$~meV are well reproduced by a
model of disordered stripes with a temperature-dependent mixture of
stripe spacings.  We discuss the relevance of our results to understanding
the magnetic excitations in cuprate superconductors.
\end{abstract}



\maketitle

\section{introduction}

High-$T_c$ superconductivity in lamellar copper oxides arises when a
sufficient density of carriers is doped into a parent Mott insulator.
Upon doping, N\'{e}el order disappears but dynamic antiferromagnetic (AF)
spin correlations survive and coexist with the induced superconductivity.
Thus, the AF spin fluctuations in a doped CuO$_2$ plane are widely
believed to have a fundamental connection with the underlying mechanism of
high-$T_c$ superconductivity.\cite{kast98}  Extensive neutron scattering
measurements have revealed an intimate relationship between the
incommensurate (IC) low-energy spin fluctuations observed in
La$_{2-x}$Sr$_{x}$CuO$_{4}$ (LSCO)\cite{yosh88,birg89b,cheo91} and
the superconductivity.\cite{yama98}  On the other hand, the discovery
of evidence for cooperative spin and charge order in
La$_{1.6-x}$Nd$_{0.4}$Sr$_x$CuO$_4$ (LNSCO) provides a new perspective on
the charge distribution within the CuO$_2$ planes~\cite{tran95a};
doped charge spatially segregates into stripes that separate antiphase
AF domains.  Such self-organized states of the strongly correlated
electrons result in a variety of interesting phenomena, and have attracted
much attention due to their potential role in the mechanism of high-$T_c$
superconductivity.\cite{emer97,cast97,vojt99,oren00,zaan01,neto01,%
carl03,arri03}

The physics behind the IC spin fluctuations in LSCO remains
controversial.  To us, the concept of fluctuating
stripes\cite{zaan96a,kive03} provides an appealing explanation of the
magnetic fluctuations; however, there is an alternative school of thought
that argues for an explanation in terms of Fermi-surface-nesting
effects.\cite{si93,litt93,manc98,kuro99,kao00,yama00,yuan01}  This
controversy is tied to the issue of whether charge-stripe order is
incompatible with superconductivity.  It is clear experimentally that
static ordering of charge stripes is correlated with a depression of
$T_c$,\cite{tran97a,ichi00} but are the excitations of the stripe-ordered
state different in nature from those in a state without static stripe
order?

To address these issues, we present a neutron scattering study of
La$_{2-x}$Ba$_x$CuO$_{4}$ (LBCO) with $x=\frac18$.  This is the material
in which high-temperature superconductivity was first
discovered\cite{bedn86} and in which the anomalous suppression of $T_c$
at $x=\frac18$ was first observed.\cite{mood88,kuma88,taka89b,kuma94b}
The difference between the Ba- and Sr-doped systems is associated with a
subtle transition from the usual low-temperature-orthorhombic (LTO)
structure of LSCO to the low-temperature-tetragonal (LTT) phase in the
Ba-doped material.\cite{axe89}  The connection between the structural
transition and the appearance of charge and spin stripe order has been
clearly demonstrated in the  $\frac18$-doped
La$_{1.875}$Ba$_{0.125-x}$Sr$_x$CuO$_4$ (LBSCO) system\cite{fuji02};
however, up until now, the occurrence of stripe order in pure LBCO
has not been confirmed due to the difficulty of growing a crystal at
the $x=\frac18$ composition.\cite{tana98,adac01}

We begin our paper by presenting neutron diffraction evidence for
stripe order within the LTT phase of La$_{1.875}$Ba$_{0.125}$CuO$_{4}$,
obtained using a large single crystal grown at Kyoto University.  On
cooling, the transition to the LTT phase begins at $T_{d2}=60$~K, and the
magnetic and charge-order superlattice peaks appear essentially
simultaneously at $T_{st} = 50$~K.  We then turn to the central topic,
which is an investigation of the spin fluctuations for excitation
energies in the range $2\le\hbar\omega\le12$~meV.  We show that these
low-energy excitations, which have some characteristics of spin waves
within the stripe-ordered
phase,\cite{zaan96a,hass99,kane01,bati01,varl02,krug03,carl04} evolve
continuously through the LTT-to-LTO transition.  For
$\hbar\omega\sim10$~meV, there is relatively little change in the
imaginary part of the dynamic spin susceptibility,
$\chi''({\bf Q},\omega)$, through the transition, while at lower energies
the {\bf Q}-integrated $\chi''$ changes from being independent of $\omega$
in the ordered state to decreasing linearly towards zero as
$\omega\rightarrow0$, as it must in the disordered state.  These changes
are similar to those observed for spin waves in undoped La$_2$CuO$_4$ as
one warms through the N\'eel temperature.\cite{yama89}  In the latter
case, the dynamical spin correlations in the paramagnetic state are
viewed as evidence of instantaneous spin correlations with the character
of the N\'eel state but without the static order.\cite{shir87,chak88}  By
analogy, we take the low-energy IC spin fluctuations in the LTO phase of
LBCO to be evidence of instantaneous stripe correlations of the same type
that become ordered in the LTT phase.\cite{kive03}  Given that the nuclear
displacements induced by charge order represent a primary order parameter
for the stripe-ordered state,\cite{tran95a,tran96b} we conclude that IC
spin fluctuations in the LTO phase are evidence of dynamic charge
stripes.\cite{kive03}  The relevance to other cuprates will be discussed.

A surprising feature observed in the ordered state is a dispersion of the
inelastic IC scattering towards the AF wave vector with increasing
$\omega$.  This is different from the behavior that is observed for spin
waves in stripe-ordered
La$_{2-x}$Sr$_x$NiO$_{4+\delta}$.\cite{tran97c,lee02,bour03,boot03}  This
mystery has been resolved in a separate study,\cite{tran04} where we
measured the spin excitations up to $\sim200$~meV and found that their
dispersion is incompatible with semiclassical spin-wave
theory.\cite{kane01,bati01,varl02,krug03,carl04}  Instead, it appears
that the full spectrum can be understood in terms of a model of weakly
coupled 2-leg AF spin ladders.\cite{vojt04,uhri04,dalo00}

Besides dispersing with energy, the apparent incommensurability $\delta$
is temperature dependent.  There is a sharp drop in $\delta$ at the
LTT-LTO transition, and then a more gradual decrease with increasing
temperature. (A similar result has been observed recently in
LNSCO.\cite{ito03})  Complementary reciprocal-space maps of the magnetic
scattering at $\hbar\omega=4$ meV have been obtained at several
temperatures for a sample of
La$_{1.875}$Ba$_{0.075}$Sr$_{0.050}$CuO$_{4}$.  We show that the latter
results can be reproduced by a disordered-stripe model, with a
temperature-dependent average stripe period.  These results may be of
interest for interpreting the charge ordering effects observed in
Bi$_2$Sr$_2$CaCu$_2$O$_{8+\delta}$ by scanning tunneling
spectroscopy (STS).\cite{howa03b,vers04,hoff02,kive03}

The format of this paper is as follows.  Sample preparation and
experimental details are described in Section II. The results of neutron
scattering measurements and the simulations based on the disordered
stripe model which reproduce the temperature dependences of observed
inelastic signal will be presented in Sections III and IV, respectively.
Then, we discuss the results and their relevance to understanding other
cuprate superconductors in Section V. Finally, our results are briefly
summarized in Section VI.

\section{Sample Preparation and experimental details}

Sizable single crystals of La$_{1.875}$Ba$_{0.125}$CuO$_{4}$ and
La$_{1.875}$Ba$_{0.075}$Sr$_{0.050}$CuO$_{4}$ were grown by a
traveling-solvent floating-zone method.  The feed rod was prepared by the
conventional solid state method.  Dried powders of La$_2$O$_3$, BaCO$_3$,
SrCO$_3$, and CuO (99.99\%\ purity) were mixed with the nominal molar
ratio of La:(Ba,Sr):Cu=1.875:0.125:1 and calcined at 860, 920 and
960$^{\circ}$C for 24 h in air with intermediate grindings.
After this procedure, we added extra BaCO$_3$ and CuO of 0.5 and 1.5
mol\%, respectively, into calcined powder in order to compensate the loss
of these components during the following crystal growth.  Mixed powder was
formed into cylindrical rods 8 mm in diameter and 150 mm in length.
The rods were hydrostatically pressed and then sintered at
1250$^{\circ}$C for 24 h in air.  We used a solvent with a composition of
La:Ba:Cu=1:1:4 (typically 350 mg in weight) and a
La$_{1.88}$Sr$_{0.12}$CuO$_{4}$ single crystal ($\sim$8 mm in diameter
and 20 mm in length) as a seed rod.

The crystal growth was performed using an infrared radiation furnace (NEC
Machinery Co., SC-N35HD) equipped with two large focusing mirrors and
small halogen lamps.  This combination of mirrors and lamps yields a sharp
temperature gradient around the molten zone, which helps to provide stable
conditions during the growth.\cite{lee98}  Both the feed and seed
rods were rotating (20/25 rpm and counter to one another) to ensure
efficient mixing of the liquid and a homogeneous temperature distribution
in the molten zone.  We set the growth rate at 0.7 mm/h and flowed oxygen
gas with the flow-rate of 100 cm$^3$/min.  These growth conditions are
essentially the same as those used for the growth of LSCO
crystals.\cite{fuji02c}  In due time, we successfully obtained a 100
mm-long crystal rod.  The initial part of the grown rods ($\sim$60 mm for
LBCO and $\sim$30 mm for LBSCO), however, broke into powder, possibly due
to the inclusion of an impurity phase such as La$_2$O$_3$ and/or
(La,Ba)$_2$Cu$_2$O$_5$.  The samples for magnetic susceptibility and
neutron scattering measurements were cut from the final part of the grown
rod.  Crystals were subsequently annealed to minimize oxygen deficiencies
in the same manner used for LSCO.\cite{fuji02c}

\begin{figure}[t]
\centerline{
\epsfxsize=2.75in\epsfbox{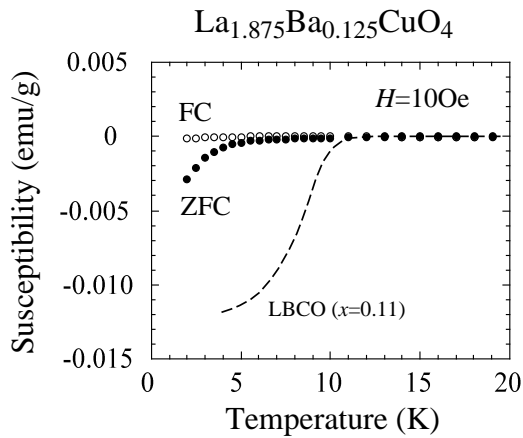}}
\caption
{
Zero-field-cooled (ZFC) and field-cooled (FC) susceptibility of
La$_{1.875}$Ba$_{0.125}$CuO$_{4}$ single crystal measured at 10 Oe. ZFC
susceptibility of La$_{1.89}$Ba$_{0.11}$CuO$_{4}$, taken from
Ref.~\protect\onlinecite{adac01}, is shown as a reference.
}
\end{figure}

Figure 1 shows the magnetic susceptibilities measured using a SQUID
(superconducting quantum interference device) magnetometer.
In the LBCO sample, the field-shielding effect at low temperature is
smaller, and the $T_c$ of 4 K is lower, compared with those reported for
La$_{1.89}$Ba$_{0.11}$CuO$_4$.~\cite{adac01}  (Here, $T_c$ is defined as
the temperature where the extrapolated slope of the low-temperature
susceptibility intersects zero.)  These results demonstrate the
suppression of superconductivity in the present LBCO crystal.
For further sample characterization, we determined the structural
transition temperatures by neutron diffraction.
With decreasing temperature, the crystal structure successively changes
from high-temperature-tetragonal (HTT, $I4/mmm$ symmetry) to
low-temperature-orthorhombic (LTO, $Bmab$ symmetry) phase at
201 K, and LTO to low-temperature-tetragonal (LTT, $P4_2/ncm$
symmetry) phase at 60 K [Fig.~4(a)], consistent with
results obtained previously on powder samples.\cite{axe89,bill93}
The former transition temperature is especially sensitive to the Ba
concentration; therefore, these results indicate that the Ba concentration
in the present sample is approximately the same as the nominal
concentration.  
The newly grown La$_{1.875}$Ba$_{0.075}$Sr$_{0.050}$CuO$_4$ sample
($T_c=9$ K) shows the LTT-LTO phase transitions at 37 K, with spin and
charge order disappearing with the structural transition upon warming.
These results are identical to those for the sample used in our earlier
elastic neutron scattering study.\cite{fuji02}

\begin{figure}[t]
\centerline{
\epsfxsize=2.25in\epsfbox{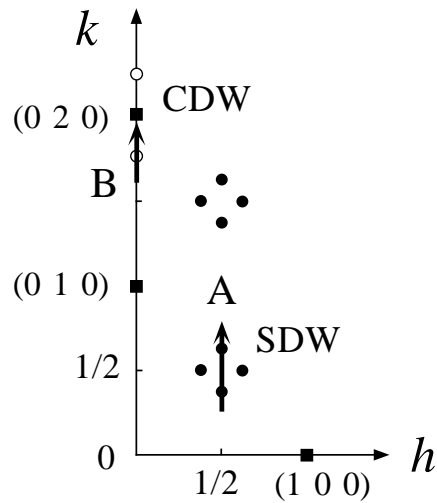}}
\caption
{
Scan geometry in the $(hk0)$ tetragonal plane. Closed squares show nuclear
Bragg peaks; open and closed circles denote nuclear and magnetic IC
superlattice peaks, respectively. }
\end{figure}

Neutron scattering measurements were performed on the Tohoku University
triple-axis spectrometer, TOPAN, installed at the JRR-3M reactor in the
Japan Atomic Energy Research Institute (JAERI).  We selected final neutron
energies $E_f$ of 14.7 meV with the collimator sequences of
$15'(30')$-$30'$-$30'$-$180'$, and 13.5 meV with
$50'$-$100'$-$60'$-$180'$, for elastic and inelastic measurements,
respectively.  Additionally, pyrolytic graphite and sapphire filters were
placed in the beam in order to eliminate higher-order neutrons.
The columnar-shaped LBCO crystal ($\sim$8 mm in diameter and 20 mm in
length) was mounted with the $(hk0)$ zone parallel to the
scattering plane.  The measurements were performed below 200 K using a
$^{4}$He-closed cycle refrigerator.
The crystal of La$_{1.875}$Ba$_{0.075}$Sr$_{0.05}$CuO$_4$ ($\sim$8 mm in
diameter and 15 mm in length) was studied on the thermal-guide
triple-axis spectrometer IN22, equipped with a double-focusing analyzer,
at the Institut Laue Langevin.  For those measurements, we
used no collimators and a PG filter was placed after the sample, with
$E_f=14.7$ meV.

In this paper, since the crystal structure of both samples at low
temperature is LTT, with in-plane lattice constant of 3.78~\AA\ (4 K), we
denote the crystallographic indices by the tetragonal notation
(1 rlu$=1.66$ \AA$^{-1}$).  Most of the inelastic scans for
La$_{1.875}$Ba$_{0.125}$CuO$_4$ were done along ${\bf Q}=(0.5,k,0)$
(denoted as scan-A in Fig.~2), which corresponds to a direction
perpendicular to the spin and charge stripes.  Therefore, the profiles
are expected to provide information on the stripe periodicity and
correlation length.

\section{Results for LBCO}
\subsection{Static correlations}

\begin{figure}[t]
\centerline{
\epsfxsize=2.4in\epsfbox{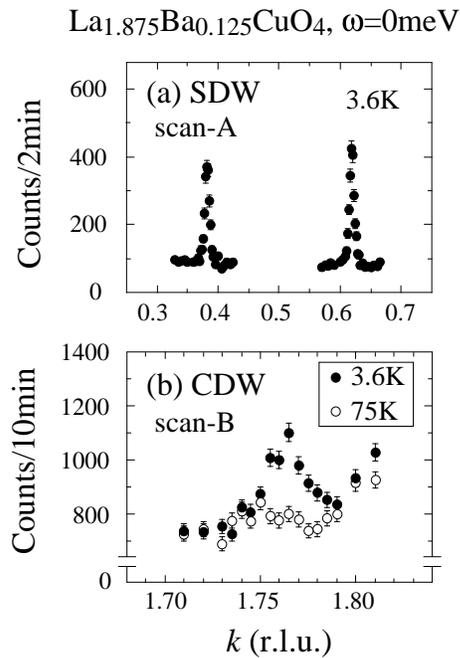}}
\caption
{
IC peaks from (a) SDW order (measured along the path labelled A in Fig. 2)
and (b) CDW order (measured along path B) in
La$_{1.875}$Ba$_{0.125}$CuO$_{4}$.  Closed (open) circles indicate
measurements below (above) $T_{st}$.  }
\end{figure}

Before investigating the spin fluctuations, we first characterize the
static stripe order in the LBCO sample with $x$=1/8.
As shown in Fig.~3, both SDW and CDW superlattice peaks were observed at
low temperature, consistent with the observations for LNSCO and
LBSCO.\cite{tran95a,tran96b,tran97a,ichi00,fuji02,fuji02b,%
niem99,vonz98,kimu03}
Both the SDW and CDW peak-widths are resolution limited, corresponding to
correlation lengths $\xi_{m}\geq150$~\AA\ for the magnetic correlations
and $\xi_{ch}\geq60$~\AA\ for the lattice modulations.
We note that the SDW and CDW peaks are found to be located at
highly-symmetric positions of $(0.5\pm\delta,0.5,0)$/$(0.5,0.5\pm\delta,
0)$ and $(2\pm\epsilon,0,0)$, respectively, where $\delta=0.118$ and
$\epsilon=0.236=2\delta$.
Therefore, the SDW and CDW wave vectors are parallel/perpendicular to the
Cu-O bond directions, as found for the tetragonal phases of
La$_{1.48}$Nd$_{0.4}$Sr$_{0.12}$CuO$_{4}$\cite{tran95a,ichi00,tran97a}
and La$_{1.875}$Ba$_{0.075}$Sr$_{0.050}$CuO$_{4}$.\cite{fuji02b}

\begin{figure}[t]
\centerline{
\epsfxsize=2.45in\epsfbox{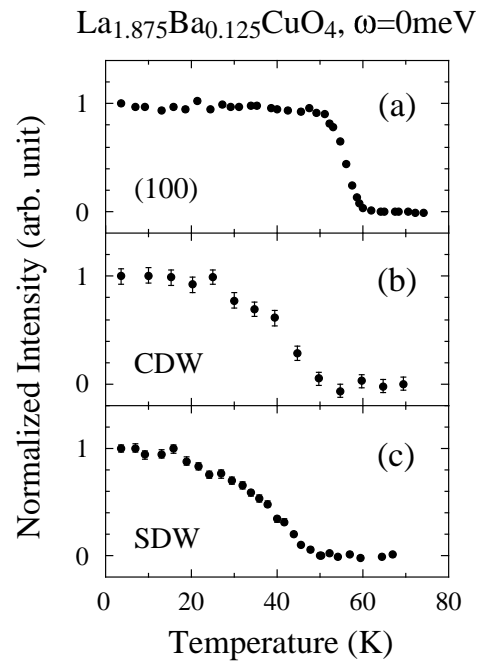}}
\caption
{
Temperature dependencies of (a) (100), (b) CDW and (c) SDW superlattice
peak intensities in La$_{1.875}$Ba$_{0.125}$CuO$_{4}$. }
\end{figure}

Upon heating, the intensity of the (100) superlattice peak associated
with the LTT structure decreases rapidly above 50~K and disappears at
$T_{d2}=60$~K because of the structural change into the LTO phase
[Fig.~4(a)].  On the other hand, both the CDW and SDW order parameters
exhibit second-order-like behavior, and the peak intensities
simultaneously vanish at $T_{st}$=50 K [Figs.~4(b) and (c)]
Coincident behavior of the two order parameters is similar to the case of
LBSCO but different from the result for LNSCO, where the SDW order
first disappears followed by the disappearance of CDW order just below
{\it T}$_{d2}$ upon heating.\cite{tran95a,tran96b,ichi00}
In contrast with the onset of the SDW and CDW orders triggered immediately
by the LTT structure in LBSCO,\cite{fuji02,fuji02d} however, $T_{st}$ is
obviously lower than $T_{d2}$ in the present sample.
The apparently simultaneous onset of magnetic and charge order in
La$_{1.875}$Ba$_{0.125}$CuO$_4$ indicates the strong correlation between
these two types of order; however, we note that muon-spin-rotation
measurements\cite{luke91} on a polycrystalline sample suggest that true
static magnetic order occurs only below 32~K.

\subsection{Dynamical correlations}

Next, we focus our attention on the spin fluctuations.
Figure 5 shows the constant-energy spectra for $\hbar\omega=3$~meV
measured at various temperatures.  In the stripe-ordered phase at 30 K,
the inelastic signal is peaked at the same wave vectors as in the elastic
scan, and the width is  resolution-limited width.
With increasing temperature, the distance between the pair of IC
peaks narrows and the peak-width grows.

\begin{figure}[t]
\centerline{
\epsfxsize=2.3in\epsfbox{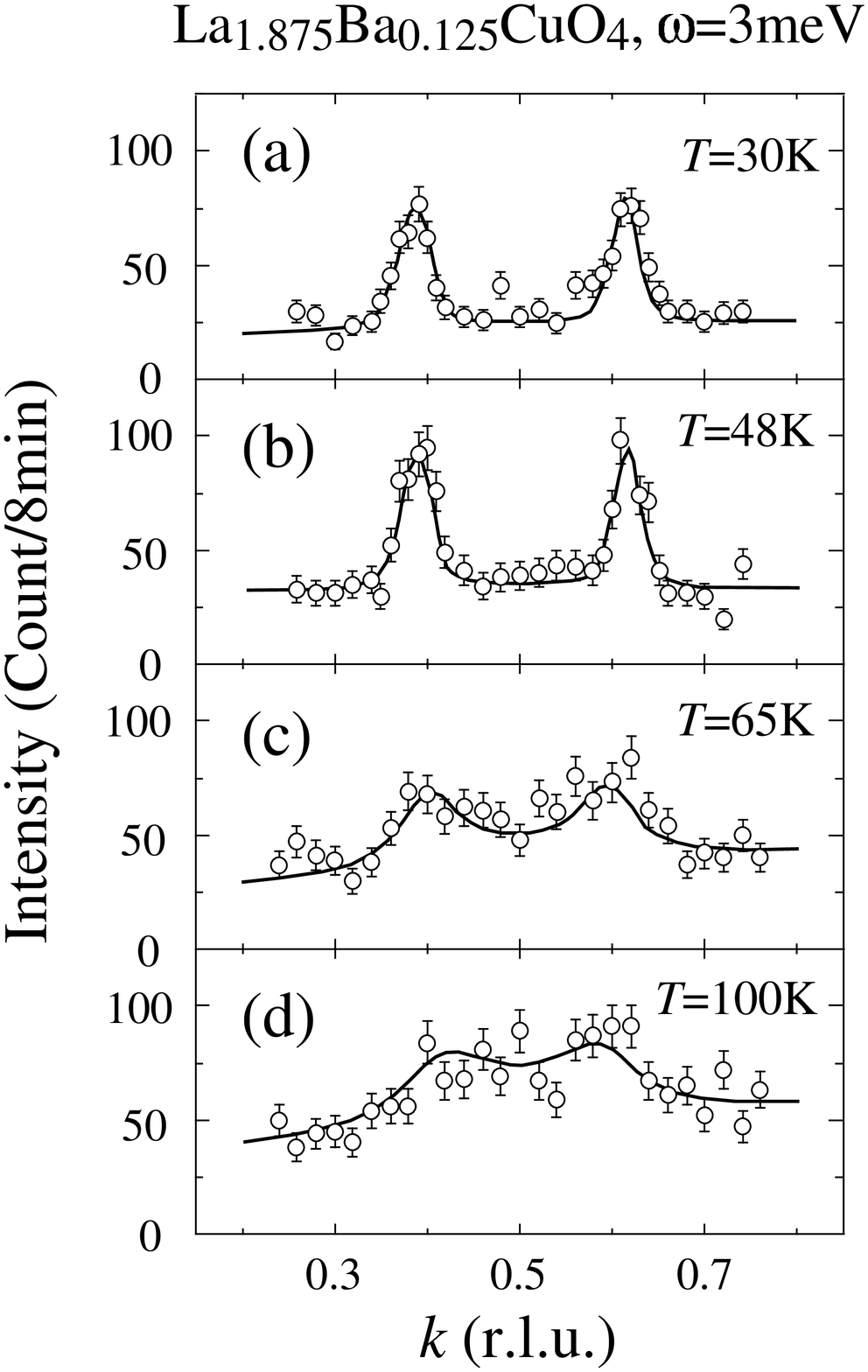}}
\caption
{
Inelastic neutron scattering spectra of La$_{1.875}$Ba$_{0.125}$CuO$_{4}$
at (a) 30 K, (b) 65 K, (c) 100 K and (d) 200 K with the constant energy
of 3 meV. The solid lines are fitted results with assuming four
equivalent peaks at $(0.5\pm\delta,0.5,0)$ and $(0.5,0.5\pm\delta,0)$.
    }
\end{figure}

Figure 6 shows a similar series of scans measured at an excitation energy
of 6 meV.  Again, sharp IC peaks are observed at $(0.5,0.5\pm0.118,0)$ in
the stripe ordered phase, while the peaks broaden and appear to merge
with increasing temperature.
Note that the IC peaks measured with $\hbar\omega=6$~meV remain
reasonably well-defined at 100 K, while the 3-meV scan yields something
closer to a single broad peak at this temperature.
We note that the $Q$-resolution at $\omega$=3 meV and 6 meV is
comparable.  Thus, the lower-energy IC spin fluctuations more easily lose
their coherence in the disordered state.

\begin{figure}[t]
\centerline{
\epsfxsize=2.3in\epsfbox{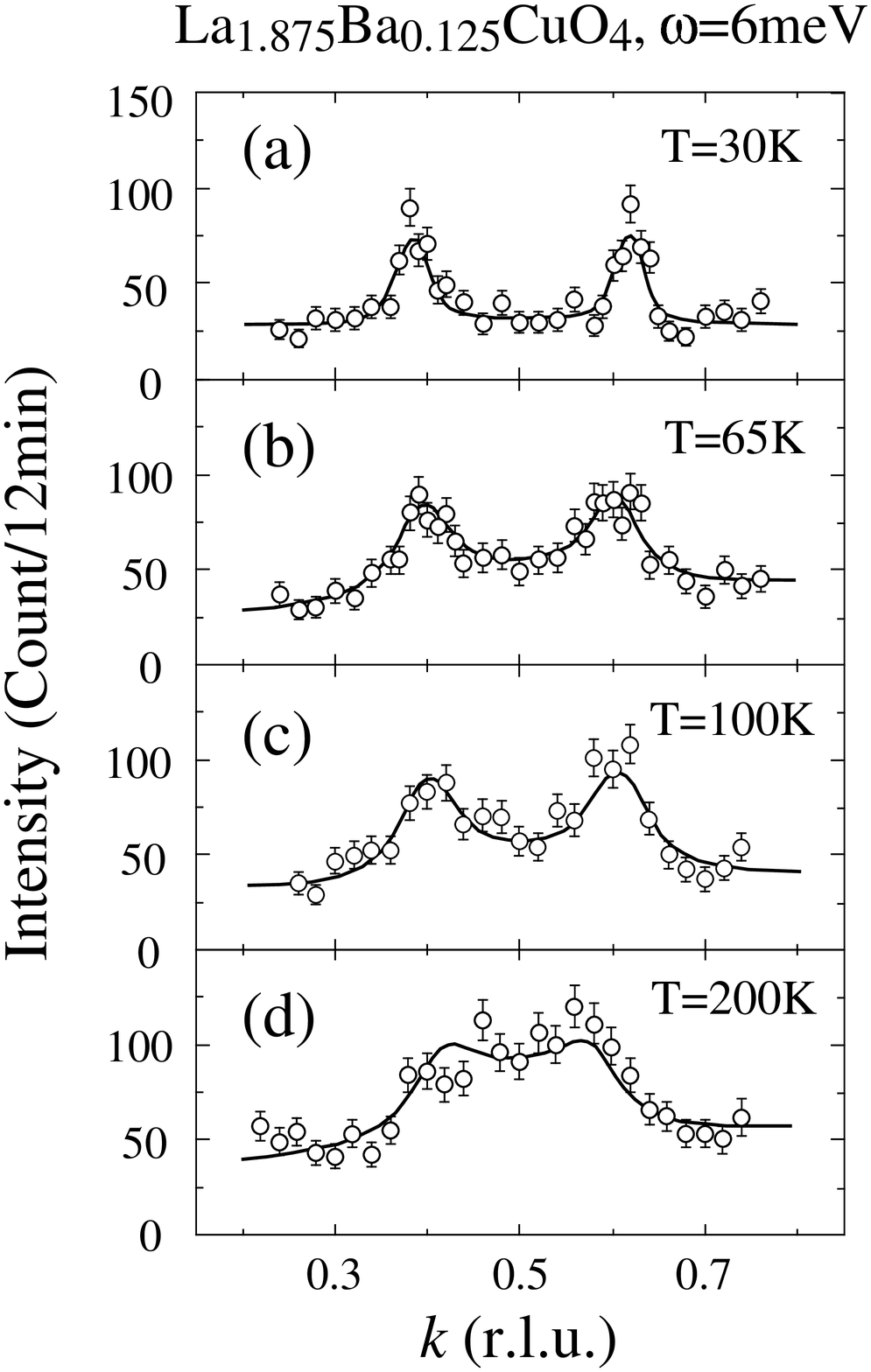}}
\caption
{
Inelastic neutron scattering spectra of La$_{1.875}$Ba$_{0.125}$CuO$_{4}$
at (a) 30 K, (b) 48 K, (c) 65 K and (d) 100 K with the constant energy of
6 meV. The solid lines are fitted results with assuming four equivalent
peaks at $(0.5\pm\delta,0.5,0)$ and $(0.5,0.5\pm\delta,0)$.
   }
\end{figure}

For quantitative analysis, we assume that the magnetic excitations
consist of four rods running along the $c^*$-axis and parameterize
$\chi''({\bf Q},\omega)$, which is proportional to the magnetic
cross-section via
$S({\bf Q},\omega) = (1-e^{-\hbar\omega/k_{B}T})^{-1}
\chi''({\bf Q},\omega)$, as follows:
\begin{equation}
    \chi''({\bf Q},\omega) = \chi''(\omega) \sum_{n=1}^{4} \frac{\kappa}
    {({\bf Q}-{\bf Q}_{\delta,n})^{2}+\kappa^2},
\end{equation}
where ${\bf Q}_{\delta,n}$ represents the four IC wave vectors,
$(0.5\pm\delta,0.5,0)$/$(0.5,0.5\pm\delta,0)$;
$\kappa$ is the peak half-width at half-maximum; and
$\chi''$ is proportional to the integral of
$\chi''({\bf Q},\omega)$ over {\bf Q} in the $(hk0)$ scattering
plane.  Measured spectra are fitted to the above function while taking
into account the experimental resolution and a background linear in $k$.

Figure 7 shows the frequency dependence of $\chi''(\omega)$ for a number
of temperatures.  In the stripe-ordered state ($T=8$ K and 30 K), it is
independent of $\omega$, just as one would expect for spin waves.  In
going from 30~K (below $T_{st}$) to 65~K (just above $T_{d2}$), there is
little change in $\chi''$ for $\hbar\omega\gtrsim8$~meV, but there is a
linear decrease towards zero at lower frequencies.  At higher
temperatures, there is a gradual reduction in the overall scale of
$\chi''$.  The modest changes observed between 30~K and 65~K indicate
that there is no significant change in the nature of the fluctuations
between the ordered and disordered states. The linear variation of
$\chi''$ with $\omega$ for low frequency at 65~K is what one expects to
see for spin fluctuations in a disordered spin system.  It is also quite
similar to what is observed in the normal state of
La$_{1.85}$Sr$_{0.15}$CuO$_4$.\cite{lee00}

\begin{figure}[t]
\centerline{
\epsfxsize=2.35in\epsfbox{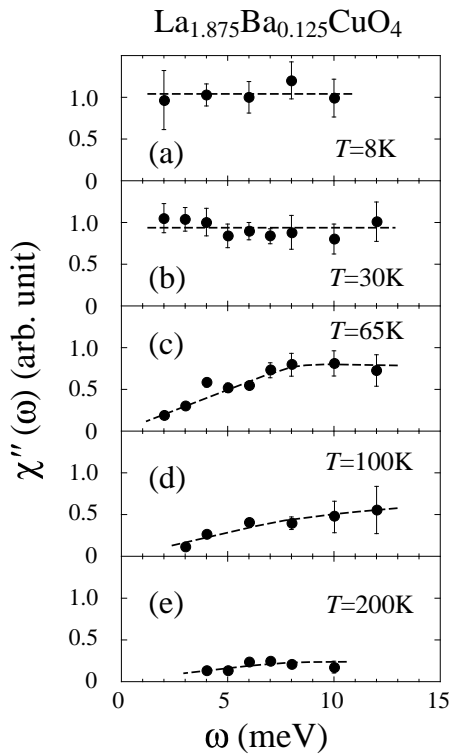}}
\caption
{
Local spin susceptibility as a function of $\omega$ in
La$_{1.875}$Ba$_{0.125}$CuO$_{4}$ at (a) 8 K, (b) 30 K, (c) 65 K, (d)
100 K, and (e) 200 K. Dashed lines are guides to the eye.  }
\end{figure}

Figure 8 summarizes the results for $\chi''$, $\kappa$ and the
incommensurability $\delta$ as a function of temperature for
$\hbar\omega=3$ and 6~meV.  The temperature dependences of all parameters
exhibit a sharp kink at $T_{d2}$, rather than at
$T_{st}$, although $\chi''$ starts to decrease at $T_{st}$
upon warming due to the disappearance of magnetic order.  The changes are
clearly larger at the smaller energy, where one is more sensitive to the
proximity to static order.  The jump in incommensurability at $T_{d2}$
suggests a lock-in effect, with the stripe spacing adjusting to be
commensurate with the modulated lattice potential that pins the stripes
in the LTT phase.\cite{hass02}  The general decrease in $\delta$ with
increasing temperature was also seen in a recent study of $\frac18$-doped
LNSCO.\cite{ito03}

\begin{figure}[t]
\centerline{
\epsfxsize=2.55in\epsfbox{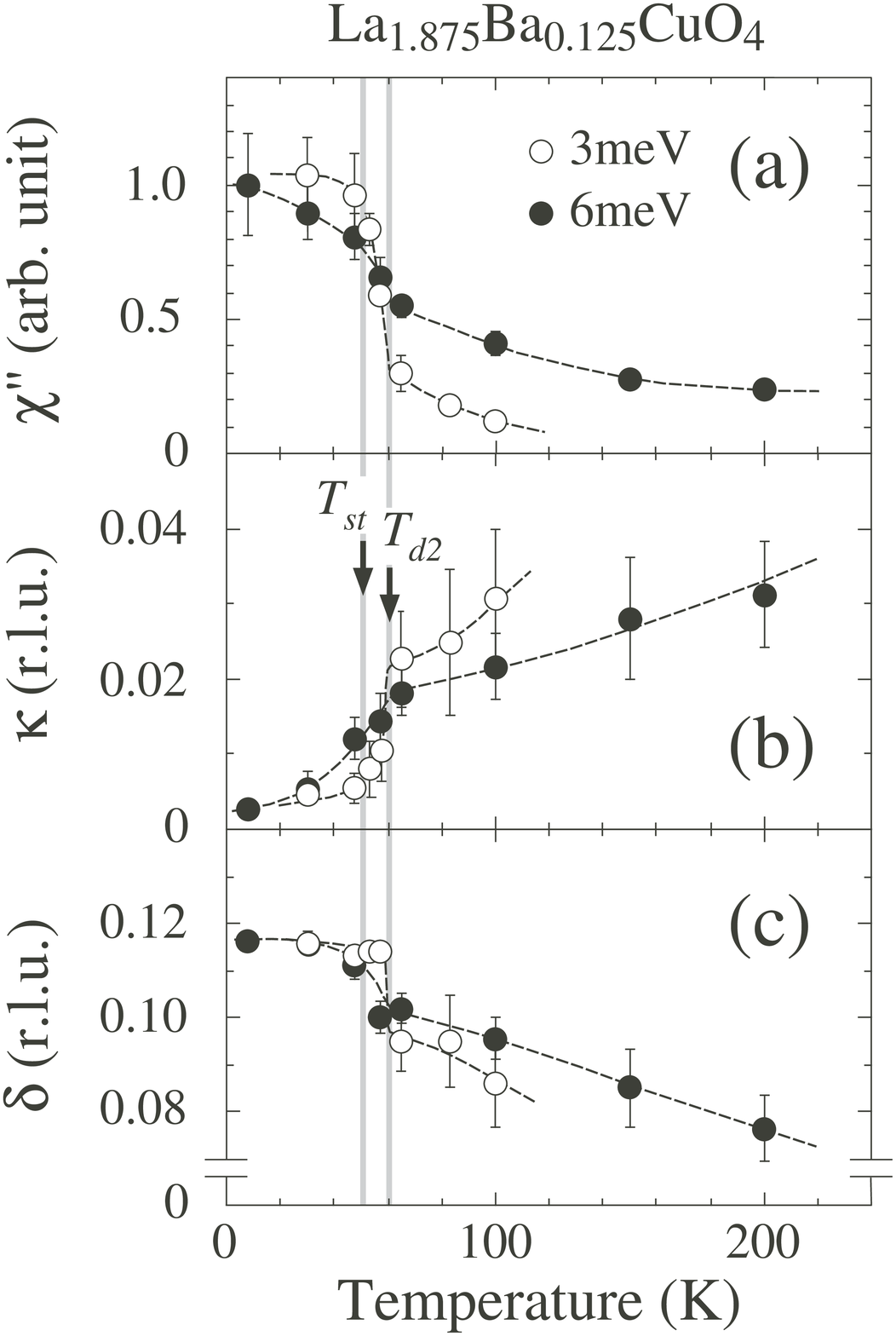}}
\caption
{
Temperature dependencies of (a) local spin susceptibility
$\chi^{\prime\prime}$, (b) peak-width in half-width at half-maximum
$\kappa$ and (c) incommensurability $\delta$ at the energy transfers of 3
and 6 meV in La$_{1.875}$Ba$_{0.125}$CuO$_{4}$. Vertical lines
indicate $T_{st}$ and $T_{d2}$. Dashed lines are guides to the eye. }
\end{figure}

Figure 9 shows the frequency-dependence of $\delta$ and $\kappa$
evaluated at 30 K, 65 K and 200 K.  At 30 K, in the stripe ordered phase,
$\delta$ gradually decreases with increasing $\omega$.
Even in the low-energy region, $\delta$ is slightly smaller than the
value of 0.125 expected from the linear relation between the hole-density
and $\delta$,\cite{yama98,tran97a} possibly due to the meandering of
stripes and/or disorder in the stripe spacing\cite{zach00,tran99a}.
Above $T_{d2}$ there is a systematic shift in $\delta$ for all $\omega$.
The dispersion of $\delta$ appears to have disappeared by the time one
reaches 200~K.

The peak half-width, $\kappa$, shows different behavior.  At 30~K, in the
stripe-ordered state, $\kappa$ increases roughly linearly in frequency.
This behavior might result from unresolved dispersion of
counter-propagating spin-wave modes.  On warming into the disordered
state at 65~K, $\kappa$ grows by a large amount at low frequencies, but
changes relatively little for $\hbar\omega\gtrsim8$~meV.  Now, this
measurement is just along a direction perpendicular to the stripes.  To
check for anisotropy, we also measured the {\bf Q}-width of the inelastic
scattering for $\hbar\omega=4$~meV at 30 K and 65 K for a direction
parallel to the stripes.  At 30 K the peak widths are isotropic within
experimental uncertainty; however, at 65 K the width perpendicular to the
stripes is roughly twice as large as that parallel to the stripes.  Such
an anisotropy might result from fluctuations in the stripe spacing.  Time
restrictions prevented a more comprehensive investigation of the
peak-width anisotropy.

\begin{figure}[t]
\centerline{
\epsfxsize=2.75in\epsfbox{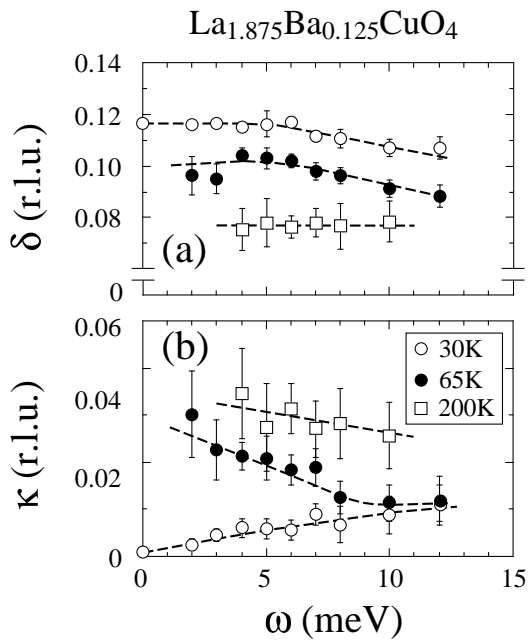}}
\caption
{
$\omega$-dependence of (a) incommensurability $\delta$ and (b) resolution
corrected peak-width (half width at half maximum) $\kappa$ of IC peaks
for La$_{1.875}$Ba$_{0.125}$CuO$_{4}$. Open circles denote 30-K data;
filled circles, 65 K; open squares, 200 K.   }
\end{figure}

\section {Disordered Stripes in LBSCO}

\subsection{Experimental measurements}

In studying the La$_{1.875}$Ba$_{0.075}$Sr$_{0.050}$CuO$_{4}$ crystal, we
performed mesh scans at an excitation energy of 4 meV, mapping out the
magnetic scattering in the neighborhood of ${\bf Q}_{\rm AF}=(0.5,0.5,0)$
for several temperatures.  All of the measurements were in the LTO phase,
where there is no static stripe order.  To present the results, it is
convenient to change to the orthorhombic coordinate system (the system in
which the mesh scans were performed), which is rotated by 45$^{\circ}$
from the tetragonal one, with a change in the lattice parameter to $a_o =
\sqrt{2}a_{t}$. In this rotated system,
${\bf Q}_{\rm AF}$ becomes (1,0,0).  The data are shown in Fig.~10(a,c,e);
a temperature-independent  background, monotonically varying in {\bf Q},
has been subtracted, and the intensities have been corrected for the {\bf
Q} dependence of the Cu$^{2+}$ magnetic form factor.\cite{sham93}
In order to improve the counting statistics, we have assumed 4-fold
symmetry of the data about ${\bf Q}_{\rm AF}$, and have averaged the data
over the corresponding rotations and reflections to give Fig.~10(b,d,e).
(The spectrometer resolution used was somewhat coarse, which reduced the
data collection time but masked any anisotropy in the peak widths at
40~K.)  As one can see, the four peaks shift in towards
${\bf Q}_{\rm AF}$ on warming, eventually merging by 200 K.

\begin{figure}[t]
\centerline{
\epsfxsize=3.33in\epsfbox{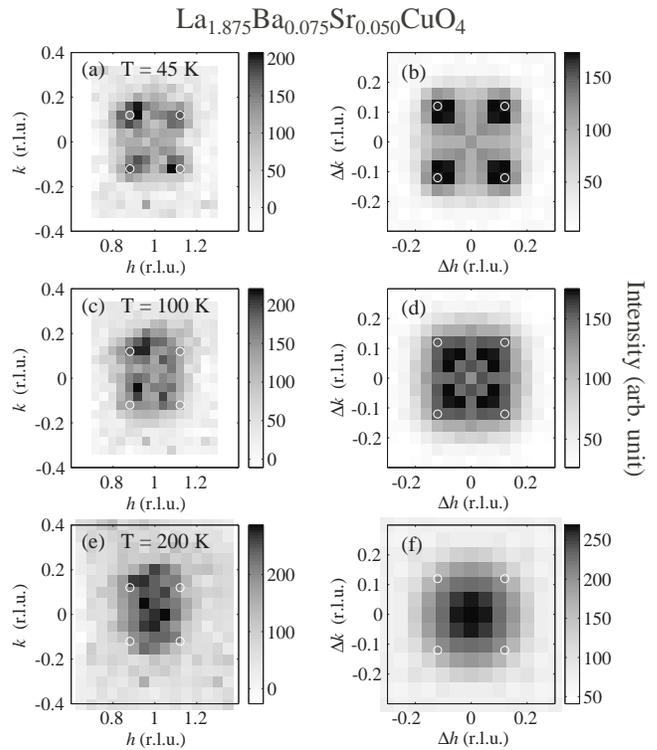}}
\caption
{
Mesh scans about ${\bf Q}_{\rm AF}$ for
$\hbar\omega$=4 meV.  (a),(c),(e) are raw data, after subtraction of
background  (see text) and correction for the Cu magnetic form factor, at
$T=45$, 100, and 200 K, respectively; panels on the right [(b),(d),(f)]
were obtained from those on the left by averaging over 4-fold symmetry
operations. The white circles indicate the positions of the elastic
magnetic peaks that appear below 37~K. }
\end{figure}

\subsection{Model calculations}

Given the shifts in the IC magnetic peaks with temperature, we want to
test how well the measurements can be described within a stripe model,
and what the data tell us about the nature of the stripe correlations.
We will assume that the {\bf Q} dependence of the low-energy fluctuations
reflects the correlations within an instantaneous configuration of
disordered stripes.  One source of disorder comes from the positions of
the charge stripes that define the magnetic domains.\cite{tran99a}  Given
a particular instantaneous configuration of stripes, we also expect there
to be a finite spin-spin correlation length.  To combine these two types
of disorder, we performed numerical calculations.  (We have also
considered transverse fluctuations in the stripe positions, but found
that the level of agreement with the measurements was not sensitive to
this additional form of disorder, so we neglect it here.)

The numerical calculations were performed on an array of
$128\times128$ sites.  The site {\bf n} was assumed to have either an up
or down spin, denoted by $F_{\bf n}=+1$ or
$-1$, or a hole, denoted by 0. Stripes of holes were taken to be straight
lines of unit width running in the $y$ direction. The spacings between the
stripes were randomly selected to be $j$ or $j+1$ (where $j$ is an
integer) with frequencies of the two choices set to give an average
spacing $d$ such that $j\le d\le j+1$.  For a given configuration, the
scattered intensity
$I({\bf Q})$ was calculated as
\begin{equation}
     I({\bf Q}) = \sum_{\bf m}
     \left(\sum_{\bf n}F_{\bf n}F^\ast_{{\bf n}+{\bf m}}
     \right) e^{i{\bf Q}\cdot{\bf m}} e^{-\kappa_s|{\bf m}|},
\end{equation}
where the exponential decay factor is intended to describe the spatial
fall off of spin-spin correlations.  The calculated intensity contains
just a pair of peaks, since the model  has a unique stripe orientation.
To compare with the measurements, we have rotated the intensity pattern
by 90$^\circ$ and added it to the original version.

In simulating the measurements, we have not made any correction for the
spectrometer resolution, which dominates the $Q$-width of the signal at
45 K.  The finite resolution, which is convolved with the sample
scattering in  reciprocal space, effectively acts like another
correlation decay factor in real space; in ignoring the resolution, we
compensate by overestimating the inverse correlation length $\kappa_s$.
The simulations are shown in Fig.~11.  At each temperature, the
parameters $d$ and $\kappa_s$ were determined by a least-squares fit to
the data along the line ${\bf Q} = (1+h,h,0)$; the comparison of the data
and simulations along this line are shown in Fig.~11(b,d,f).  The
parameter values for each temperature are listed in Table 1.  The
disordered stripe model appears to give an adequate description of the
data.

\begin{figure}[t]
\centerline{
\epsfxsize=3.25in\epsfbox{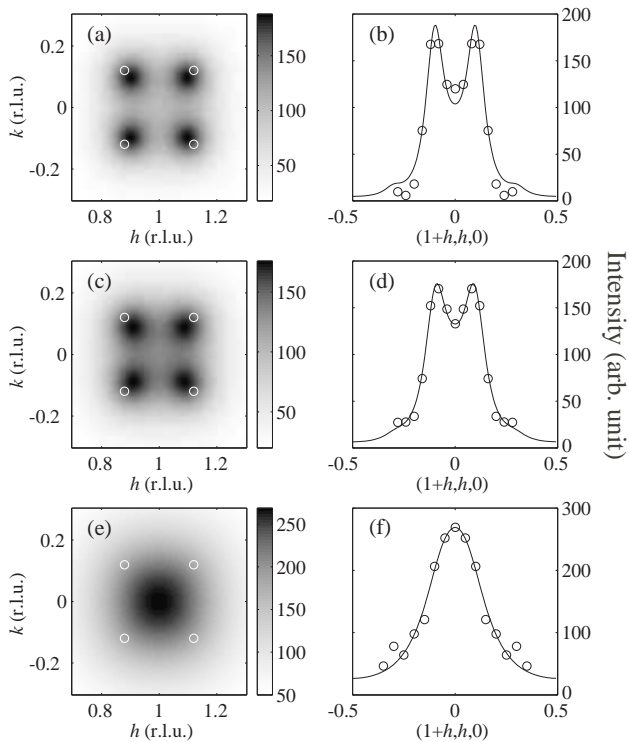}}
\caption
{
Simulations of the 4-meV scans, as described in the text.  (a), (c),
and (e) are simulations of the mesh scans at $T=45$, 100, and 200~K,
respectively. (b), (d), and (f) compare the calculated curves with the
symmetrized data along ${\bf Q} = (1+h,h,0)$. The white circles are the
same as in Fig.~10.
}
\end{figure}

\begin{table}
\caption{Parameter values determined by fitting simulations to data. $d$
is expressed in lattice units and
$\kappa$ in r.l.u., both for the tetragonal cell.}
\begin{ruledtabular}
\begin{tabular}{rcc}
$T$ & $d$ & $\kappa_s$ \\
(K) & & (r.l.u.) \\
\hline
45 & $5.0\pm0.4$ & $0.06\pm0.01$\\
100 & $5.2\pm0.4$ & $0.07\pm0.01$ \\
200 & $7.1\pm1.4$ & $0.14\pm0.03$ \\
\end{tabular}
\end{ruledtabular}
\end{table}

One key result of the modelling is that one must allow for a significant
change in the stripe spacing from that in the ordered state in order to
get a reasonable fit to the data.  A second important result is that the
scattering becomes commensurate as the correlation length
($=1/(2\pi\kappa_s)$ in lattice units) becomes smaller than half of the
stripe spacing.  Thus, instantaneous charge stripe correlations are
compatible with commensurate magnetic fluctuations.

The values of the average stripe spacing required to describe the
measurements in the disordered state are significantly longer than the
value of $\approx4$ that is characteristic of the ordered state.  How can
we understand this?  Transverse fluctuations of a given stripe increase
its arc length.\cite{kive98}  If the hole density per arc length remains
roughly the same as in the ordered state, then there must be more holes
within each stripe.  That, in turn, implies a lower density of stripes
within a CuO$_2$ plane, and hence an increased stripe spacing.  Repulsive
Coulomb interactions between stripes would also tend to favor increased
average stripe spacings when transverse fluctuations are important.  The
fact that adequate simulations of the scattering measurements do not
require explicit inclusion of the transverse meanderings may simply
indicate that the spin-spin correlation length is shorter than the
typical distance between transverse stripe displacements.

\section{discussion}

Our results indicate that the spin correlations in the LTO phase of
La$_{1.875}$Ba$_{0.125}$CuO$_{4}$ are a dynamic form of the ordered state
found at lower temperatures in the LTT phase.  The low-temperature phase
is characterized by charge-stripe order.  While magnetic order appears to
have the same onset temperature (50 K), a muon-spin-rotation study has
shown that the true static order appears only below 30 K.\cite{luke91}
  From the perspective of doped antiferromagnets, stripe order involves a
spatial segregation of doped holes that allows a persistence of
hole-poor, AF-insulator regions.  We have obtained direct evidence for
the existence of a dynamic, fluctuating stripe phase in LBCO.  A related
dynamic stripe phase has been detected previously in
La$_{2-x}$Sr$_x$NiO$_4$,\cite{lee02,bour03} so there is a precedent for
such behavior.

The abrupt change in the incommensurability at $T_{d2}$ may allow an
improved understanding of the striking behavior of the Hall coefficient
measured by Noda, Eisaki, and Uchida\cite{noda99} in
La$_{1.4-x}$Nd$_{0.6}$Sr$_x$CuO$_4$.  For that system, the Hall
coefficient behaved ``normally'' (i.e., looked similar to comparably doped
LSCO) in the LTO phase, but dropped rapidly towards zero on cooling
through the transition to the LTT phase for $x\le0.13$.  The behavior in
the LTT phase has been explained in terms of the response of charge
stripes with a doped-hole concentration of 0.5 per Cu site.  For such a
condition, there is electron-hole symmetry within a stripe, and
consequently the Hall coefficient should be zero.\cite{emer00,prel01}
The electron-hole symmetry is quite sensitive to the hole concentration
in the stripes, and hence it should be sensitive to the
incommensurability, which will affect the hole concentration.  The jump
in the incommensurability that we observe at the structural transition
implies a jump in the hole concentration within the stripes, assuming
that all holes remain in stripes.  (Even if the average hole
concentration within stripes remains roughly the same due to meandering
of the stripes in the disordered phase, there may still be enough change
to eliminate the particle-hole symmetry.)  It is reasonable to expect
similar behavior in La$_{1.4-x}$Nd$_{0.6}$Sr$_x$CuO$_4$, so that the
abrupt change in the Hall coefficient likely reflects the difference in
hole concentration for dynamic stripes versus that in static stripes.

Recent analyses of the Hubbard model using methods beyond Hartree-Fock
have yielded stripe solutions that are quite similar to experimental
observations.\cite{lore02,loui01,flec00,whit03,ichi99b}  A
first-principles calculation for $\frac18$-doped LSCO using the LDA+U
method yields bond-centered charge stripes\cite{anis04}; the superexchange
interactions calculated within the hole-poor ladders are comparable to
what we have obtained from measurements of the magnetic excitations in
LBCO at higher energies.\cite{tran04}  Thus, there is growing theoretical
support for the concept of stripe correlations as a natural consequence of
doping holes into an antiferromagnetic insulator.  Quantum fluctuations
(and the absence of a pinning potential) lead to the stripe-liquid
state.\cite{zaan96a,kive98,sach99,momo03,schm03}

\subsection{Relevance to LSCO}

The low-energy magnetic fluctuations found in the normal state of
LSCO\cite{lee00} look quantitatively similar to what we have measured in
the LTO phase of LBCO.  Elastic IC magnetic peaks can be induced in
underdoped LSCO at low temperature through Zn-doping\cite{kimu03b} or by
applying a magnetic field along the $c$-axis.\cite{kata00,lake02}  The
simplest explanation for all of the these observations is that dynamic
charge stripes are present in LSCO and that they can be pinned by local
defects.\cite{waki03}

Weaknesses in the Fermi-surface-nesting explanation for the IC spin
fluctuations in LSCO have been discussed by Kivelson {\it et
al.}\cite{kive03}  Given the experimental evidence for the dynamic stripe
phase presented here, we believe that the Fermi-surface-nesting approach
is no longer tenable for interpreting results in LSCO.  Rather than
trying to explain the IC spin correlations in terms of the shape of the
Fermi surface, one must strive to understand photoemission measurements of
the electronic spectral function near the Fermi surface\cite{zhou04} in
terms of the slowly fluctuating charge stripes.

The gapless spin fluctuations in the normal state of LSCO indicate that
the charge stripes must fluctuate quite slowly.  This raises the question
of whether there is some feature of LSCO that might control the
fluctuation rate.  An old, but still plausible, idea is that the charge
stripes may couple to the octahedral tilt mode that is associated with
the transformation from the LTO to the LTT phase.  This mode has an
energy of just a couple of meV.\cite{brad94}  It softens on cooling below
100 K, but the softening ends at $T_c$.\cite{lee96,kimu00}  A hardening
of the elastic constant $(C_{11}-C_{12})/2$ below $T_c$ was found to be
reduced by the lowering of $T_c$ through application of a magnetic
field.\cite{noha93}  Thus, it seems quite possible that charge stripes in
LSCO are coupled to slow LTT-like fluctuations of the lattice.

\subsection{Relevance to YBCO}

There has long been a recognition of similarities in the low-energy
magnetic scattering of well-underdoped YBCO with that of
LSCO,\cite{tran92,ster94} and measurements to higher energies made clear
similarities to antiferromagnetic dispersions.\cite{bour97}  The clear
identification of incommensurate magnetic scattering at $\sim24$ meV in
YBa$_2$Cu$_3$O$_{6.6}$ by Mook {\it et al.}\cite{mook98,dai98,mook00} made
the connection to LSCO stronger.  Recent studies have provided strong
evidence for stripe-like spin excitations in detwinned crystals of
YBa$_2$Cu$_3$O$_{6.5}$,\cite{stoc04} and for both charge and spin
modulations in YBa$_2$Cu$_3$O$_{6.35}$.\cite{mook02}

Objections to the dynamic stripe picture have come from studies of spin
excitations in YBCO samples closer to optimal doping.\cite{bour00}  There
the excitations observed in the superconducting state are incompatible
with semiclassical spin waves from
stripes.\cite{kane01,bati01,varl02,krug03,carl04}  An interpretation of
these features based on Fermi-surface-nesting effects has been
preferred.\cite{bour00,ito02,kao00}  Our study of the high-energy spin
excitations in LBCO removes the objection to a stripe interpretation, as
the results are quite different from the predictions of semiclassical
spin-wave models.  In fact, our results for LBCO show striking
similarities to recent measurements on YBCO samples with a range of
dopings.\cite{arai99,rezn03,hayd04}  Thus, it appears that the
dynamic-stripe scenario provides a universal approach for understanding
most features of the magnetic excitation spectrum in the two most
carefully studied systems, YBCO and LSCO.  The magnetic resonance
phenomenon is one feature that is not yet explained by this approach.

There is one apparent difference between LSCO and YBCO that can be
explained by the dispersion of the low-energy spin excitations.  In LSCO,
the incommensurability $\delta$, measured at $\sim3$~meV, varies linearly
with doping up to a hole concentration $p\approx\frac18$, and it
saturates at $\delta\sim\frac18$ for larger $p$.\cite{yama98}  In YBCO
there is a substantial spin gap that grows with $p$, and hence one must
measure $\delta$ at relatively high energies ($>30$ meV near optimum
doping).  Dai {\it et al.}\cite{dai01} found that $\delta$ saturates at
$\sim\frac1{10}$ for $p>0.10$, which is different from the LSCO result.
Taking the dispersion of the spin excitations into account should
reduce this apparent discrepancy.

\subsection{Relevance to STS studies}

Scanning tunneling spectroscopy studies on
Bi$_2$Sr$_2$CaCu$_2$O$_{8+\delta}$ have identified spatial modulations of
electronic states at low energies (within the
superconducting gap).\cite{howa03b,vers04,hoff02,kive03}  Some of the
spatially-modulated features disperse with bias voltage, and these appear
to be associated with scattering of electronic excitations across the
gapped Fermi surface.\cite{hoff02,vers04}  Certain features, however,
involve modulations oriented parallel to the Cu-O bonds with a period of
approximately $4a$, suggesting a connection with the type of
charge-stripe order that we have discussed
here.\cite{howa03b,vers04,kive03,podo03}  The relevant STS modulations
do not always have a period of exactly $4a$.  The period varies slightly
from sample to sample, and may increase a bit with temperature.  Here we
point out that such behavior is quite consistent with the
temperature-dependent incommensurability in LBCO and with the doping
dependent incommensurability in LSCO.\cite{yama98}

\section{summary}

We have presented a neutron scattering study of stripe order and
fluctuations in single crystals of La$_{1.875}$Ba$_{0.125}$CuO$_4$ and
La$_{1.875}$Ba$_{0.075}$Sr$_{0.050}$CuO$_{4}$.  Charge and spin stripe
order are observed only within the LTT phase.  The {\bf Q}-integrated
dynamic susceptibility is frequency independent in the ordered state,
consistent with spin waves; however, the spin excitations disperse
inwards towards ${\bf Q}_{\rm AF}$ with increasing energy in an
anisotropic manner that is not expected in a semiclassical model.

The IC spin excitations evolve continuously through the LTT-LTO
transition.  For $\hbar\omega\sim10$~meV, there is essentially no change
in the local susceptibility through the transition, indicating that the
character of the excitations in the disordered state is the same as in
the ordered state, where the spin incommensurability is tied to the
presence of charge stripes.  Our measurements provide clear evidence for
dynamic charge stripes in the LTO phase.  We have discussed the relevance
of our results for interpreting the magnetic excitations observed in LSCO
and YBCO.

\section*{Acknowledgements}

We thank H. Kimura, K. Machida, M. Matsuda, G. Shirane, H. Yamase, I.
Watanabe, and G. Xu for valuable discussions. This work was supported in
part by the Japanese Ministry of Education, Culture, Sports, Science and
Technology, Grant-in-Aid for Scientific Research on Priority Areas (Novel
Quantum Phenomena in Transition Metal Oxides), 12046239, 2000, for
Scientific Research (A), 10304026, 2000, for Encouragement of Young
Scientists, 13740216, 2001 and for Creative Scientific Research (13NP0201)
``Collaboratory on Electron Correlations - Toward a New Research Network
between Physics and Chemistry'', by the Japan Science and Technology
Corporation, the Core Research for Evolutional Science and Technology
Project (CREST).  JMT is supported at Brookhaven by the Office of
Science, U.S. Department of Energy, through Contract No.\
DE-AC02-98CH10886.


\end{document}